\documentclass[10pt]{article}
\usepackage{amssymb}
\usepackage{amstext}

\newcommand*\de{\mathrm{d}}
\newcommand*\De{\mathrm{D}} 
\renewcommand*\epsilon{\varepsilon}
\renewcommand*\phi{\varphi}
\renewcommand*\theta{\vartheta}

\newcommand*\AL{{}^{\text{\tiny L}}\!\! A}
\newcommand*\EL{{}^{\text{\tiny L}}\!\! E}
\newcommand*\EE{{}^{\text{\tiny T}}\!\! E}
\renewcommand*\AA{{}^{\text{\tiny T}}\!\! A}
\renewcommand*\P{{}^{\text{\tiny T}}\!\pi}
\newcommand*\PP{{}^{\text{\tiny TT}}\!\pi}
\newcommand*\HH{{}^{\text{\tiny TT} }\!h}
\newcommand*\HV{{}^{\text{\tiny V} }\!h}
\newcommand*\PV{{}^{\text{\tiny V} }\!\pi}
\renewcommand*\H{{}^{\text{\tiny T}}\!h}

\begin{document}
\title{\bf \large Gauge 
invariance of the wave functional in mixed \\
momentum/coordinate representations}

\author{ M. Leclerc\\ \small
  Section of Astrophysics and Astronomy,
Department of Physics,\\ \small  University of Athens, Greece}
 \date{\small March 8, 2007}
\maketitle
\begin{abstract}
Starting from the observation that in Yang-Mills theory 
the Schroedinger state functional in the momentum representation is not 
gauge invariant, we investigate the reversed question: Which are the 
representations for the operators of a gauge theory that lead to an 
invariant wave functional 
once the quantum constraints have been imposed upon it? Stated otherwise:  
Which representation do we have to use if we wish the constraints of the 
theory to eliminate the non-physical degrees of freedom from the states? 
We use the framework of geometric quantization to attack this question. 
In particular, it is found that in the linear spin-two theory  
as well as in 
General Relativity, gauge invariance cannot be achieved by a pure 
coordinate (i.e., field)  representation,  but that one has to use 
 mixed momentum/coordinate representations instead. Our results are 
illustrated by the example of the free relativistic point-particle as well as 
by simple cosmological mini-superspace models in the framework of 
General Relativity.  
\end{abstract}

\section{Introduction}

The momentum representation is well known as an alternative, but 
equivalent representation in classical Schroedinger 
quantum mechanics. Therefore, at first sight, the results 
presented in \cite{jackiw} and \cite{jackiw2} might appear surprising. 
It is shown for  non-abelian Yang-Mills theory that  in the conventional 
coordinate (i.e., field) representation, the state functional is 
gauge invariant, while this is not the case 
in the momentum representation. Such an  asymmetry between coordinates 
and momenta is certainly not in the 
spirit of the Hamiltonian formulation. One is then led to the question 
whether this is a general feature of gauge theories, or one might 
speculate that there is  
 something special about the coordinate representation. This is not the 
case though, and it turns out that the above asymmetry is actually less 
deep than  it appears. Indeed, as far as the physical, unconstrained 
variables are concerned, there is a complete symmetry between coordinates 
and momenta, just as we know it from ordinary quantum mechanics, while 
an asymmetry occurs only as a result of the constraints and therefore 
affects only  the constrained and the gauge variables, but not the 
dynamical ones.    

While this is not hard to recognize, it does still not answer our 
question. Is the coordinate wave functional generically 
gauge invariant? More precisely, 
what one wishes to do, in a gauge theory, is to use a representation 
where $X$ acts on the states by multiplication 
($X$ represents collectively the fields (or the coordinates) of the theory)
and the momenta by differentiation, 
$\Pi = \frac{1}{i}\frac{\delta }{\delta X}$. Then one applies the constraint 
operator $\hat G(X,\Pi)$ on the states $\psi(X)$. The solutions 
of the equations $\hat G(X,\Pi) \psi(X) = 0$  are then expected to 
be gauge invariant, i.e., to depend only on the gauge invariant 
components of $X$. In this way, {\it the constraints 
remove the unphysical degrees of freedom  from the physical states of the
theory}, a statement that can be found in almost any textbook on quantum 
field theory. As we have said before, this is true in the Yang-Mills case 
(and obviously  in electrodynamics). It turns out, however, that it is not 
always true. Consider, for instance, the fourth constraint in 
the linear spin-two theory, $h^{ik}_{\ \ ,i,k} - \Delta h = 0$. In the 
quantum theory, this leads to the condition 
$(h^{ik}_{\ \ ,i,k} - \Delta h) \psi (h_{ik}) = 0 $ on the 
state functional. But in the coordinate representation where $h_{ik}$ acts 
by multiplication, this cannot be seen as condition on $\psi$ (since else, 
we have to conclude $\psi = 0$), but rather on $h_{ik}$, i.e., we have to 
solve the constraint classically such that $h^{ik}_{\ \ ,i,k} - \Delta h = 0$, 
and then formulate the quantum theory with the remaining components of
$h_{ik}$. This is what one does during the Hamiltonian reduction 
of the theory, see \cite{faddeev}, but it is not what we have in mind here. 
Here, we wish the constraint itself, as a quantum operator,  
to eliminate the non-physical variables from 
the state functional.  This is obviously not the case for the above 
constraint. It is the case for the remaining three constraints of the 
spin-two theory, $\pi^{ik}_{\ \ ,i}= 0$,  
but summing up, we have to conclude that,  
 in the coordinate representation, the 
constraints do not remove all the unphysical degrees of freedom from the 
state functional. It is rather obvious what we have to do: For those 
components of $h_{ik}$ that appear in the expression $h^{ik}_{\ \ ,i,k} 
- \Delta h$, we have to use a momentum representation and turn them into 
differentiation operators. Simultaneously then, some components of 
the momenta $\pi^{ik}$ will turn into multiplication operators. It turns out 
that we are lucky inasmuch those specific components do not appear in 
the expression $\pi^{ik}_{\ \ ,i}$, such that the remaining three constraints 
still eliminate three degrees of freedom from $\psi$. Therefore, in this 
new, mixed coordinate/momentum representation, the constraints do indeed 
what they are supposed to do, namely they remove  the unphysical 
degrees of freedom from the theory.  
 
Now we have answered our  question, namely {\it the coordinate  wave 
functional is not necessarily gauge invariant, but there might be 
an invariant  functional corresponding to a different, possibly mixed, 
 representation}. 
(By invariant functional, we always understand a functional that 
becomes invariant upon imposing the quantum constraints).  
All we need  now is a systematic way to identify the corresponding 
representation(s) for  general gauge theories. That is the main subject of 
the present article. 
A convenient approach to the problem is provided by 
the formalism of geometric quantization, the basic features of which we  
present  in the next section.  Explicit applications are presented in the 
remaining sections. 

\section{Geometric quantization}\label{geo}
  
A clear and concise introduction to the formalism of geometric quantization 
can be found in \cite{jackiw2}. The presentation provided there is completely 
sufficient for our purposes, and there is also no way to present things 
in a simpler way. For the mere purpose of self-containedness of our 
paper, we will repeat step by step the presentation of \cite{jackiw2}. 
On the occasion, we will introduce a new index notation that we find 
convenient for our later applications. 

We start from a set of canonical coordinates (or fields) 
 $q^A$ and corresponding 
momenta $p_A$, where $A$ can be any kind of index, discrete of continuous. 
By canonical, we mean that the kinetic term in the first order Lagrangian 
is (up to a total derivative) of the form $p_A \dot q^A$, resulting in 
the standard symplectic two-form (see below). For the specific subject 
we wish to study, it is important to keep track of the momenta and 
coordinates. Therefore, instead of labeling the phase-space variables 
with an index running over twice the range of the index of $q^A$,   
we use a  notation of the form 
\begin{equation} \label{1}
\xi^{A\alpha} = (p_A, q^A), \quad \alpha = 1,2.  
\end{equation}
The symplectic two-form $\omega = \frac{1}{2}(\omega_{AB})_{\alpha\beta} \de
\xi^{A\alpha}\xi^{B\beta}$ then has the components $\delta_{AB} 
\epsilon_{\alpha\beta}$, with antisymmetric $\epsilon_{\alpha\beta}$, and 
$\epsilon_{12} = 1$.  This form is exact, $\omega = \de \theta$, 
and thus, we can write 
\begin{equation} \label{2}
\omega = \de \theta = \frac{1}{2} (\omega_{AB})_{\alpha\beta} \de 
\xi^{A\alpha}\de \xi^{B\beta}, \quad 
(\omega_{AB})_{\alpha\beta} = \frac{\partial \theta_{B\beta}}{\partial 
\xi^{A\alpha}} - \frac{\partial \theta_{A\alpha}}{\partial \xi^{B\beta}}
= \delta_{AB}\  \epsilon_{\alpha\beta}.
\end{equation}
The physical object is $\omega$, while  
 $\theta = \theta_{A\alpha} \de \xi^{A\alpha}$ is only defined up to 
an exact form. 
The conventional choice is $\theta_{A\alpha} = (0, p_A)$, leading to 
$\theta = p_A \de q^A$. Note also that in the case of field theory, 
partial derivatives have to be replaced by variational derivatives
and $\delta_{AB}$ by a delta-function. 

Next, to each phase-space function 
$G(\xi)$, we introduce a corresponding vector 
field $v(\xi)$ via 
\begin{equation}\label{3}
v^{A\alpha} (\omega_{AB})_{\alpha\beta} = - \frac{\partial G}{\partial 
\xi^{B\beta}}, \quad \text{or equivalently} \quad
v^{A\alpha}  = - \epsilon^{\alpha\beta}  
\frac{\partial G}{\partial \xi^{A\beta}}. 
\end{equation}
The function  $G(\xi)$ is also 
called the generator of the vector field $v(\xi)$, 
but we will not make use of this vocabulary in order to avoid confusion 
with the generators of symmetry transformations. Further, for the same 
quantity $G(\xi)$, we introduce an operator $\hat G$ defined as 
\begin{equation}\label{4}
\hat G = \frac{1}{i} v^{A\alpha} \De_{A\alpha} + G, \quad \text{with} \quad
\De_{A\alpha} = \frac{\partial}{\partial \xi^{A\alpha}} - i \theta_{A\alpha}. 
\end{equation}
Those operators, referred to as pre-quantized operators, 
 act on the so-called pre-quantized wave functions 
(or functionals) $f(\xi)$ that vary over the entire phase space, 
depending thus on both $q^A$ and $p_A$. 

In particular, for the coordinates $q^A = \xi^{A2}$, we find 
from (\ref{3}) the vector $v^{AB\alpha} = (- \delta^{AB}, 0)$ (there is 
an additional index on $v$ here, because $G = q^A$ already carries an index), 
and from (\ref{4}) the operator 
\begin{equation} \label{5}
\hat q^A = \frac{1}{i}\left[ 
- \frac{\partial}{\partial p_A} + i \theta^A_{\ 1}\right] 
+ q^A. 
\end{equation}
Similarly, for $p_A = \xi^{A1}$, we have $v^{AB\alpha} = (0,\delta^{AB})$
and
\begin{equation} \label{6}
\hat p_A = \frac{1}{i}\left[  \frac{\partial}{\partial q^A} - i \theta_{A2}
\right]+ p_A. 
\end{equation}
The explicit form depends on the choice of $\theta$. 

The final step towards the quantum theory is the choice of a so-called 
polarization. This consists in choosing vector fields $\pi$ which 
span half the phase-space and imposing the following 
conditions on the pre-quantized wave functions $f(\xi)$
\begin{equation}\label{7}
\pi^{BA\alpha}\De_{A\alpha} f(\xi) = 0. 
\end{equation}

For instance, choosing for $\pi^{AB\alpha}$ the vector generated by 
the coordinates $q^A = \xi^{A2}$, i.e., $\pi^{AB\alpha} = (- \delta^{AB},0)$, 
the above condition takes the form 
\begin{equation} \label{8} 
\De_{A1} f = (\frac{\partial f}{\partial \xi^{A1}} - i \theta_{A1})f = 0, 
\end{equation}
which leads, with the conventional choice $\theta_{A\alpha} = (0,p_A)$ 
to $\frac{\partial f}{\partial p_A} = 0$, meaning that the 
function $f(\xi)$ depends merely on the coordinates and not on the 
momenta, i.e., $f = \psi(q^A)$. (In addition, $f$ may also depend on time, 
but we suppress this in our notation.) 
This corresponds to the coordinate representation. Indeed, it is
straight-forward to verify that (\ref{5}) and (\ref{6}) now reduce 
to the conventional representations $\hat q^A = q^A$ and 
$\hat p_A = \frac{1}{i} \frac{\partial }{\partial q^A}$. In the same 
way, the momentum representation is obtained by choosing for 
$\pi^{AB\alpha}$ the vector generated by the momenta $p_A$, see
\cite{jackiw2}.  We will illustrate this for the example of 
non-abelian Yang-Mills theory at the end of this section. 

Finally, we turn to the study of symmetry generators. Classically, 
such a generator $G$ induces a change in the phase-space 
variables $\xi$ via 
\begin{equation}
[G, \xi^{A\alpha}] = 
\frac{\partial G}{\partial \xi^{B\beta}} \ 
[\xi^{B\beta}, \xi^{A\alpha}] 
= \frac{\partial G}{\partial \xi^{B\beta}}\ (\omega^{BA})^{\beta\alpha} 
\equiv - \delta \xi^{A\alpha}, 
\end{equation}
where $[{\ } ,{\ } ]$ is the canonical 
Poisson bracket and $(\omega^{AB})^{\alpha\beta}= - \delta^{AB}
\epsilon^{\alpha\beta}$ is the inverse of $(\omega_{AB})_{\alpha\beta}$.  
 In components, we find 
\begin{equation}\label{9}
\delta q^A = \frac{\delta G}{\delta p_A}, \quad \delta p_A = - \frac{\partial
G}{\partial q^A},  
\end{equation}
which leads, via (\ref{3}) to the vector field $v^{A\alpha} = 
(\delta p_A, \delta q^A) = \delta \xi^{A\alpha}$. (This explains the 
earlier mentioned vocabulary, namely to call $G$ the generator 
of the corresponding vector field $v$. Here, however, we are interested 
in those specific $G$'s for which $\delta \xi$ is indeed a symmetry of 
the theory.) With $v$ determined, we can construct the pre-quantized 
operator corresponding to $G$ and we find 
\begin{equation} \label{10} 
\hat G = \frac{1}{i} \left[\delta p_A \frac{\partial }{\partial p_A} 
+ \delta q^A \frac{\partial }{\partial q^A} \right]
- \left[\delta p_A \theta^A_{\ 1} + \delta q^A  \theta_{A2} - G\right]. 
\end{equation}
Acting on the pre-quantized function $f(q^A, p_A)$, we get 
\begin{equation} \label{11}
\hat G f = \frac{1}{i} \delta f - 
\left[\delta p_A \theta^A_{\ 1} + \delta q^A  \theta_{A2} - G\right]f 
= \frac{1}{i} \delta f - 
\left[\delta \xi^{A\alpha} \theta_{A\alpha} - G\right] f,  
\end{equation}
with $\delta f = \frac{\partial f}{\partial q^A}
 \delta q^A + \frac{\partial f}{\partial p_A} \delta p_A$.  This is 
the relation that will be used in all of the following, because it 
states that, if  $\hat G$ is a symmetry operator, then  
the constraint $\hat G f = 0$ on the wave function leads to 
$\delta f  = 0$ (i.e., $f$ gauge invariant) only if    
$\delta \xi^{A\alpha} \theta_{A\alpha} - G = 0$. Thus, if this is the case, 
and if we in addition, we use the corresponding representation 
for which the wave 
function $\psi$ of the quantum theory is equal to the 
pre-quantized wave function   
after the choice of the polarization (as was the case for the previously 
presented coordinate representation, $f = \psi(q^A)$), then we can conclude 
that the action of $\hat G$ on $\psi$ is equivalent to a gauge transformation, 
and consequently, imposing the constraint $\hat G \psi = 0$ eliminates 
the unphysical (gauge) degrees of freedom from the theory. 

Since there was a double {\it if} in the above argumentation, 
one might ask whether it is possible for both {\it if}'s to be 
violated, and the conclusion to be nevertheless true. In other words, 
can it happen that  $f$ is not gauge 
invariant, and that $f $ is not equal to the quantum wave function $\psi$, 
and that $\psi$ is nevertheless gauge invariant? This is indeed possible, 
and is related to the fact that the one-form $\theta$, that appears 
explicitely in (\ref{11}), is not uniquely determined, since only the 
two-form $\omega = \de \theta$ is invariant under canonical transformations 
and thus physically relevant. Therefore, the question whether $\hat G f = 0$ 
leads to $\delta f = 0$ or not cannot be  of direct physical relevance, 
since the answer depends on $\theta$. It turns out, however, that a change 
of $\theta$ by an exact form leads, in the quantum theory, to a 
shift in the phase of the wave function, and to a corresponding change 
in the operator representations defined by (\ref{5}) and (\ref{6}),  such 
that ultimately, the phase can be discarded and the resulting 
quantum theory is equivalent to the initial one. Instead of 
a general proof, we briefly sketch the corresponding issue on 
the example of non-abelian Yang-Mills theory. 

Using the notations of  \cite{jackiw2}, we have the momenta 
$p_A \equiv E^i_a$ and the fields $q^A \equiv A^a_i$. The 
symmetry generator is given by ($\epsilon^a$ is an arbitrary parameter)
\begin{equation}\label{12}
G = \int \de^3 x\ \epsilon^a (\partial_i E^i_a + f_{ab}^{\ \ c} A^b_i E^i_c). 
\end{equation}
We thus have $\frac{\delta G}{\delta A^b_i} = f_{ab}^{\ \ c}E^i_c 
\epsilon^a \equiv  - \delta E^i_b$ and $\frac{\partial G}{\partial E_a^i}
 = - \epsilon^a_{\ ,i} + f_{cb}^{\ \ a} A^b_i \epsilon^c \equiv 
 \delta A^a_i$. 
(The fact that we use the same symbol $\delta$ both for the 
change induced by the symmetry generator, e.g., $\delta A^a_i, \delta f$, 
etc., as well as for the functional derivatives should not lead to confusion.) 

A crucial observation is to recognize that we can write 
$G = - \int \delta A^a_i E^i_a \de^3 x$, and therefore, 
from (\ref{11}), we find 
\begin{equation}\label{13}
\hat G f(A,E) = \frac{1}{i}\delta
 f(A,E) - \int\de^3 x \left[\delta E^i_a \theta^a_{i1} + 
\delta A^a_i \theta^i_{a2} - \delta A^a_i E^i_a \right]  \ f(A,E). 
\end{equation}

First, we make  the conventional choice that has also been used in 
\cite{jackiw}, namely $\theta^a_{i1} = 0$ and $\theta^i_{a2} = E^i_a$. 
We then find immediately $\hat G f = \frac{1}{i} \delta f$, meaning 
that once the constraint $\hat G f$ is imposed, $f$ will be gauge 
invariant. In the coordinate representation, we have the 
polarization $\De_{i1}^a f = 0$, i.e., $\frac{\delta f}{\delta E^i_a} = 0$. 
Thus, the quantum functional can be chosen to be  equal to  $f$, and 
does not depend on $E^i_a$, $f = \psi(A^a_i)$. In particular, 
$\psi(A^a_i)$ is thus gauge invariant (after imposing the constraint). 
We can easily check that the operators are in the usual representation,
i.e., $\hat A^a_i = A^a_i$ and $\hat 
E^i_a = \frac{1}{i}\frac{\partial}{\partial 
A_i^a}$. 

On the other hand,  the momentum representation is obtained from 
the polarization $\De_{a2}^i f = 0$, which leads to 
$\frac{\delta f}{\delta A^a_i} - i E^i_a f = 0$, with 
 solution $f = \exp(i \int E^i_a A^a_i \de^3 x )\phi(E)$. 
Hence, we cannot directly use $f(A,E)$ as momentum representation 
wave functional, but rather, we have to use $\phi(E)$, which depends 
only on half of the phase-space variables. Since $f$ was gauge 
invariant, $\phi(E)$ is obviously not. This is the result obtained 
in \cite{jackiw} and \cite{jackiw2}. From (\ref{5}) and (\ref{6}), 
we now find the operators 
$\hat A^a_i = - \frac{1}{i} \frac{\delta}{\delta E^i_a}
+ A^a_i$ and $\hat E^i_a = E^i_a$. Applied to the above $f$, we see 
that we can eliminate the phase factor $\exp(i\int E^i_a A^a_i\de^3 x)$, 
i.e., 
we have 
$\hat A^a_i \phi(E) = - \frac{1}{i} \frac{\delta}{\delta E^i_a} \phi(E)$ 
and $\hat E^i_a \phi(E) = E^i_a \phi(E)$ which corresponds indeed 
to the conventional momentum representation.

It is not unnatural to suspect that the asymmetry between momentum and 
coordinate representations was brought in by hand, namely when we 
made the choice $\theta^a_{i1} = 0$ and $\theta^i_{a2} = E^i_a$. 
In order to show that this is not the case, and that $\theta$ can 
be chosen arbitrarily (provided only that $\de \theta = \omega$), 
we now repeat the above analysis with  the symmetric  
choice $\theta_{i1}^a = -\frac{1}{2} A^a_i $ and $\theta^i_{a2} = 
\frac{1}{2} E^i_a$. This differs from the previous once by a 
total differential and thus leads to the same symplectic two-form 
$\de \theta$. From (\ref{13}), we now find $\hat G f(A,E) = \frac{1}{i} \delta 
f(A,E)
+ \int \frac{1}{2} \delta(A^a_i E^i_a) \de^3 x f(A,E)$. We see that now, 
$f$ is not gauge invariant. Applying again the coordinate polarization 
$\De_{i1}^a f = 0$ leads to $\frac{\delta f}{\delta E^i_a} + \frac{i}{2} 
A^a_i f = 0$, with solutions $f = \exp(-\frac{i}{2}\int E^i_a A^a_i \de^3 x)
\phi(A)$. Thus, in the quantum theory, we cannot use $f(A,E)$, but 
have to use $\phi(A)$. Also, the operators $\hat A^a_i$ and $\hat E^i_a$ 
acting on $f(A,E)$ are modified, but on $\phi(A)$, they act 
again as multiplication 
and differentiating operators, respectively,  
just as they did in the original coordinate representation on $f(A)$. 
And finally, we also find  $\hat G f = \exp(-\frac{i}{2} \int E^i_a A^a_i 
\de^3 x) \delta \phi(A)$, meaning that upon imposing the constraint
$\hat G f= 0$, we find $\delta \phi = 0$
and thus $\phi(A)$ is again gauge invariant. Summarizing, the change of 
$\theta$ by an exact form results in an unphysical  phase shift, which can 
be removed since the operators are modified accordingly, and the resulting 
wave functional is again gauge invariant. In the same fashion, one shows that 
for the above $\theta$, the wave functional in the momentum representation 
is again not gauge invariant. The asymmetry between momentum and coordinates 
is thus not introduced by a specific choice of $\theta$ (which is 
physically irrelevant), but it is a fundamental property of the theory
itself. Namely, it enters the theory via the constraint.

Although the choice of $\theta$ is physically irrelevant, one would 
consider the above choice as unwise, in the sense that it only complicates 
the analysis. If we assume that for a given gauge theory, there exists  
always  a representation for which the wave functional is gauge 
invariant (after imposing the constraint), then it is obviously 
preferable to choose $\theta$ directly in a way that the 
pre-quantized functional $f(\xi)$ is already gauge invariant and then 
look for the representation for which the quantum functional $\psi$ 
is directly given by the pre-quantized functional after polarization, 
i.e., $f = \psi$.  If instead we start from a pre-quantized functional 
that is not gauge invariant, it will be hard to find the representation 
that leads to an invariant functional in the quantum theory. 

It is useful to observe that from (\ref{5}) and (\ref{6}), one can read off 
 the following correspondence 
\begin{eqnarray}
\theta^A_1 = - q^A,\  \theta_{A2} = 0 \quad \text{implies}\quad 
f = \psi(p_A) \quad \text{in the momentum polarization}\quad \De^A_{\ 2}f 
= 0, \\
\theta^A_1 = 0,\  \theta_{A2} = p_A \quad
\text{implies}\quad f = \psi(q^A) \quad \text{in the coordinate 
polarization}\quad \De_{A1} f = 0. 
\end{eqnarray}
Thus, e.g., for the previously adopted choice $\theta^A_1 = -\frac{1}{2} q^A$ 
and $\theta_{A2} = \frac{1}{2} p_A$, the quantum functional $\psi$ does not  
 directly correspond to the pre-quantized  functional $f$ in neither the 
coordinate nor the momentum representation. 

This ends our introduction to geometric quantization and we will now 
consider concrete examples. 

\section{Linear spin-two theory}\label{spin2}

The first order Lagrangian for the spin-two theory reads 
\begin{equation}\label{15}
L =\int\left [\pi^{ik} \dot h_{ik} -  H( \pi_{ik}, h_{ik}) 
+ h_{0 i} ( 2 \pi^{ik}_{\ \ ,k}) + 
h_{00}
(-\frac{1}{2} h^{ik}_{\  \ ,i,k} + \frac{1}{2} \Delta h
)\right] \de^3 x,   
\end{equation}
where our signature  convention is $\eta_{ik} =- \delta_{ik}$  
($i,k\dots = 1,2,3$), $h = h^{i}_{\ i}$ and 
$\Delta h = h^{,i}_{\ ,i} h$.
The Hamiltonian has the form 
\begin{equation} \label{16}
 H =  \int \left [\pi^{ik} \pi_{ik} - \frac{1}{2} \pi^2 
+ \frac{1}{4} 
h_{,i} h^{,i} 
- \frac{1}{4} h^{ik}_{\ \ ,m} 
h_{ik}^{\ \ ,m}+ \frac{1}{2} 
h^{ik}_{\ \ ,i} h_{mk}^{\ \ ,m} 
- \frac{1}{2} h^{ik}_{\ \ ,i}h_{,k}\right]\de^3 x. 
\end{equation}
The Lagrangian  is invariant under the transformations 
\begin{equation}\label{17}
\delta h_{ik} = \xi_{i,k} + \xi_{k,i}, \quad
\delta \pi_{ik} = - \epsilon_{,i,k} + \eta_{ik} \Delta \epsilon,
\quad
\delta h_{0i} = \dot \xi_{i} + \epsilon_{,i}, \quad 
\delta h_{00} = 2 \dot \epsilon. 
\end{equation}
The transformations of $\pi^{ik}$ and $h_{ik}$ are generated by  
\begin{equation}\label{18}
G_{\xi} = -\int \xi_i 2 \pi^{ik}_{\ \ ,k} \de^3 x,  \quad 
G_{\epsilon} = - \int \epsilon \left
[- h^{ik}_{\  \ ,i,k} +  \Delta  h  \right]\de^3 x,
\end{equation} 
corresponding to the constraint equations $\pi^{ik}_{\ \ ,k} = 0$ and 
$-h^{ik}_{\ \ ,i,k} + \Delta h = 0$. 

It is convenient to write the canonical fields $h_{ik}$ and $\pi^{ik}$ 
in the form 
\begin{eqnarray} \label{19}
h_{ik} &=& \HH_{ik} + (\eta_{ik} - \frac{\partial_i \partial_k}{\Delta}) \H 
+ h_{i,k} + h_{k,i} = \HH_{ik} + \H_{ik} + \HV_{ik} \\ \label{20}
\pi^{ik} &=& \PP^{ik} + (\eta^{ik} - \frac{\partial^i \partial^k}{\Delta}) \P 
+ \pi^{i,k} + \pi^{k,i} = \PP^{ik} + \P^{ik} + \PV^{ik}, 
\end{eqnarray}
where $\H_{ik}$ and $\P^{ik}$ are transverse and $\HH_{ik}$ and $\PP^{ik}$ 
traceless transverse, see \cite{adm} for details. The $TT$, $T$ and $V$ 
components are easily shown to be orthogonal under integration. 
Note that we now have 
\begin{eqnarray}
\delta_{\xi} h_{ik} = \delta_{\xi} \HV_{ik} = \xi_{i,k} + \xi_{k,i}, 
\quad \text{or} \quad \delta_{\xi} h^i = \xi^i \\
\delta_{\epsilon} \pi_{ik} = \delta_{\epsilon} \P_{ik} = \Delta \epsilon 
\eta_{ik} - \epsilon_{,i,k}, \quad \text{or}\quad \delta_{\epsilon}
 \P  = \Delta
\epsilon, 
\end{eqnarray}
where from now on, we denote by $\delta_{\epsilon}$ ($\delta_{\xi}$) 
the transformations induced by $G_{\epsilon}$ ($G_{\xi}$). The remaining
variations are zero, i.e., $\delta_{\epsilon} \PP_{ik} = \delta_{\epsilon}
\pi^i = \delta_{\epsilon} h_{ik} = 0$ and $\delta_{\xi} \HH_{ik} = 
\delta_{\xi} \H = \delta_{\xi} \pi_{ik} = 0$. 
The Lagrangian can now be written in the form 
\begin{equation}\label{23}
L = \int \left[\PP^{ik}\, \dot{ \HH_{ik}} + 2 \P\, \dot{ \H} 
+ 2 (\pi^{i,k} + \pi^{k,i}) 
\dot h_{i,k} 
- H + h_{0k} (2 \Delta \pi^k + 2 \pi^{i,k}_{\ \ ,i}) + h_{00}\, 
\Delta\H\right]\de^3 x, 
\end{equation}
with $H = 
\int\left[\PP_{ik}\PP^{ik} + 2 \pi_{i,k}\pi^{i,k} - 4 \P \pi^{i}_{\ ,i} 
- \frac{1}{4} \HH_{ik,l} \HH^{ik,l}\right]\de^3x$. If we wish to reduce 
the theory by solving the constraints, we find $\H= \pi_i= 0$, and 
the theory reduces to its dynamical variables $\PP^{ik}$ and $\HH_{ik}$
and can be trivially quantized,  
see \cite{adm,leclerc,leclerc2}. Here, instead, we wish to follow the 
conventional Dirac method, which consists in imposing the constraint 
on the quantum states of the theory, $\hat G_{\xi} \psi = \hat 
G_{\epsilon} \psi = 0$. 
This procedure, in electrodynamics and in Yang-Mills theory, has the 
effect of eliminating the unphysical (gauge) degrees of freedom from 
the states in the coordinate representation, $\psi =\psi(A)$ (suppressing 
again the explicit time dependence in our notation). Here, however, as 
we have already mentioned in the introduction, this is not the case. 
Imposing $\hat G_{\xi}\psi(h_{ik}) = 0$ leads to $\psi = \psi(\HH_{ik}, 
\H)$, while the fourth constraint $\hat G_{\epsilon} \psi(h_{ik})$ 
acts simply by multiplication. In other words, if we insist on 
using the coordinate   representation, then we have to solve 
the constraint $- h^{ik}_{\ \ ,i,k} + \Delta h = 2 \Delta \H 
= 0$ classically, eliminating 
thereby $\H$ {\it before} we perform the transition to the quantum theory. 
This, however, is a reduction in the sense of \cite{faddeev}, as we have 
performed it in \cite{leclerc,leclerc2}, but it is not the Dirac 
procedure.

If we wish to stick faithfully to the Dirac procedure, it is obvious what we 
have to do: The field $\H$ occurring in the fourth constraint 
has to be taken in the momentum representation. More systematically, 
using the formalism of the previous section, we are looking for 
a pre-quantized functional that is gauge invariant under both 
$\hat G_{\xi}$ and $\hat G_{\epsilon}$. According to (\ref{11}), 
we have 
\begin{eqnarray} \label{24}
\hat G_{\xi} f 
&=& \frac{1}{i} \delta_{\xi}f - \left(\int \left[ \delta_{\xi} \pi^{ik} 
\theta_{ik1} + \delta_{\xi} h_{ik} \theta^{ik}_{\ \ 2}\right]\de^3 x
-G_{\xi}\right) f
= 
\frac{1}{i} \delta_{\xi}f - 
\left(\int \left[ \delta_{\xi} h_{ik} \theta^{ik}_{\ \ 2}\right]\de^3 x
- G_{\xi} \right)f \\\label{25}
\hat G_{\epsilon} f &=& 
\frac{1}{i} \delta_{\epsilon}f - \left(\int \left[ \delta_{\epsilon} \pi^{ik} 
\theta_{ik1} + \delta_{\epsilon} h_{ik} \theta^{ik}_{\ \ 2}\right]\de^3 x
- G_{\epsilon} \right)f
= 
\frac{1}{i} \delta_{\epsilon}f - \left(\int \left[ \delta_{\epsilon} \pi^{ik} 
\theta_{ik1}\right]\de^3 x  
- G_{\epsilon}\right) f. 
\end{eqnarray}
We also note that $G_{\xi}$ is linear in $\pi^{ik}$, while $G_{\epsilon}$ is 
linear in $h_{ik}$, and therefore, we find $G_{\xi} = \int 
(\delta_{\xi} h_{ik}) 
\pi^{ik} \de^3x$ 
and $G_{\epsilon} = -\int (\delta_{\epsilon} \pi^{ik}) h_{ik} \de^3 x$. 
Therefore, it is straightforward to recognize that with the 
conventional choice $\theta_{ik1} = 0$ and $\theta^{ik}_{\ \ 2} = \pi^{ik}$ 
(that leads to $f = \psi(h_{ik})$ in the coordinate polarization), 
we find that $f$ (and thus $\psi$) is invariant under $\hat G_{\xi}$, but 
 not under $\hat G_{\epsilon}$, in accordance with our earlier observations. 
An obvious choice that makes $f$ invariant under both $G_{\epsilon}$ and 
$G_{\xi}$ is given by $\theta_{ik1}= - h_{ik}$ and $\theta^{ik}_{\ \ 2} = 
\pi^{ik}$. But this is not allowed, since it leads to the double value 
of the symplectic two-form, $\de \theta = 2 \omega$. The crucial point is 
to recognize that not all  components, e.g., of  $\theta_{ik1}$ 
occur in $\int \delta_{\epsilon} \pi^{ik} \theta_{ik1} \de^3 x$, since 
$\delta_{\epsilon} \pi^{ik}$ contains only transverse parts, and similar 
for the remaining terms in (\ref{24}) and (\ref{25}). Indeed, if 
we choose 
\begin{equation} \label{26}
\theta_{ik1} = - (\eta_{ik} - \frac{\partial_i \partial k}{\Delta}) \H = 
- \H_{ik}, \quad \theta^{ik}_{\ \ 2} = \PP^{ik}
+ \pi^{i,k} + \pi^{k,i} = \PP^{ik} + \PV^{ik} = \pi^{ik} - \P^{ik}, 
\end{equation}
then, in view of the orthogonality of the different tensor parts, 
we find that $\hat G_{\xi} f = \frac{1}{i} \delta_{\xi} f $ 
and $\hat G_{\epsilon} f = \frac{1}{i} \delta_{\epsilon} f $, and it is 
also obvious that $\de \theta$ leads to the canonical form $\omega$.   
For the existence of the above choice of $\theta$, it is crucial that 
the constraints are first class. This ensures that, if for instance 
$\P^{ik}$ occurs in one constraint, then $\H_{ik}$ does not occur 
in the other, and as a result, we have exactly the correct number of 
independent, orthogonal variables occurring in the generators to make 
$f$ invariant under all of the generators and to still guarantee that 
 $\de \theta  = \omega$. It is also clear that for the dynamical 
variables, $\PP_{ik}$ and $\HH^{ik}$, which do not appear in the 
constraints, we have the choice to put them into either component of 
$\theta_{\alpha}$, e.g., as $\PP^{ik}$ in $\theta^{ik}_{\ \ 2}$ as 
above or as $-\HH_{ik}$ into $\theta_{ik1}$ (which would lead to 
a momentum representation). Mixed representations are also allowed
for those variables. This reflects again the fact that for the 
dynamical variables, there is no asymmetry whatsoever between momenta and 
fields. The asymmetry enters the theory only via the constraints and 
therefore affects only the constrained variables $(\H, \pi^i)$ and 
the corresponding gauge variables $(\P, h_i)$. This is true for 
any gauge theory. Thus, for instance, if we wish to use a momentum 
representation in electrodynamics and still have a gauge invariant 
wave functional, then the obvious thing to do is to put the 
transverse fields $(\P^i, \AA_i)$ into the momentum representation, 
and leave the longitudinal fields in the coordinate representation.  
In the non-abelian case, the situation is more involved, since the 
physical components of $A^a_i$ are not so easily identified, 
see, e.g.,  \cite{jackiw}. 

For the transition to the quantum theory, we need to choose a polarization. 
What we are looking for is of course the polarization that leads to 
an identification between the pre-quantized functional $f$ and the 
corresponding quantum functional $\psi$. In fact, the polarization, 
and thus the variables on which $\psi$ depends,  can directly 
be read off from (\ref{26}): The {\it momenta} are those fields 
that occur in $\theta_{ik1}$ and $\theta^{ik}_{\ \ 2}$ (i.e., those 
fields should be represented with differentiation operators), while 
 the {\it coordinates} are the conjugate  fields. From (\ref{26}), 
we then find that $\psi = \psi(\HH_{ik},\HV_{ik},\P^{ik})$. In other 
words, $\psi$ depends on the gauge variables, and not on the constrained 
ones. This is obviously the only way in order for the constraints to be 
able to remove the unphysical variables. 

Somewhat more systematically, the above is achieved with the 
polarization vector 
\begin{equation}\label{27}
\pi_{lm\ \ 1}^{\ \ ik}= \frac{1}{2}\left[\delta^i_l\delta^k_m + 
\delta^k_l \delta^i_m - 
(\eta_{lm}- \frac{\partial_l \partial_m}{\Delta})
(\eta^{ik}- \frac{\partial^i \partial^k}{\Delta})
\right], \quad 
\pi^{lm}_{\ \ ik2} = \frac{1}{2}(\eta^{lm}- \frac{\partial^l\partial^m}
{\Delta})
(\eta_{ik} - \frac{\partial_i \partial_k}{\Delta}). 
\end{equation}
Imposing, according to (\ref{7}),  the polarization on the pre-quantized 
functional, 
$[\pi_{lm\ \ 1}^{\ \ ik} \De_{ik1}+ \pi_{lmik2} \De^{ik}_{\ \ 2}]f=0$, then 
leads to (the meaning of the symbolic notation is obvious)
$\left[{}^{\text{\tiny V}}\!(\De_{lm1}) + {}^{\text{\tiny TT}}\!(\De_{lm1})
+ {}^{\text{\tiny T}}\!(\De_{lm2})\right]f = 0$. Since the 
three operators are orthogonal, each term has to vanish by itself. 
Explicitely,  with 
$\theta$ from  (\ref{26}), we find $\De_{lm1} = \frac{\delta }{\delta
  \pi^{lm}} +  i\,\H_{lm}$ and $\De^{lm}_{\ \ 2} = \frac{\delta}{\delta 
h_{lm}} - i\, \PP^{lm} - i\, \PV^{lm}$, leading to the ten 
constraints 
\begin{equation}
\frac{\delta}{\delta \PV^{lm}} f = \frac{\delta}{\delta \PP^{lm}} f 
= \frac{\delta}{\delta \H_{lm}} f = 0, 
\end{equation} 
with solution  $ f = \psi(\HH_{lm}, \HV_{lm}, \P^{lm})$, or 
equivalently, $\psi(\HH_{lm}, h_i, \P)$. Finally, we construct the 
operators $\hat h_{ik} = \frac{1}{i} \De_{ik1} + h_{ik}$ and 
$\hat \pi^{ik} = \frac{1}{i} \De^{ik}_{\ \ 2} + \pi^{ik}$. Decomposing into 
orthogonal components, we find 
\begin{eqnarray}\label{29}
\hat {\HH_{ik}} &=& \HH_{ik}, \quad \hat{\HV_{ik}} = \HV_{ik}, \quad 
\hat{\H_{ik}} = - \frac{1}{i} \frac{\delta }{\delta \P^{ik}}, \\ 
\hat {\PP^{ik}} &=& \frac{1}{i} \frac{\delta }{\delta \HH_{ik}}, \quad 
\hat {\PV^{ik}}= \frac{1}{i} \frac{\delta }{\delta \HV_{ik}}, 
\quad \hat {\P_{ik}} = \P_{ik}. \label{30}
\end{eqnarray}
Imposing the constraints $\hat G_{\xi} \psi = \hat G_{\epsilon} \psi = 0 $ 
on $\psi(\HH_{ik}, \HV_{ik}, \P_{ik})$ now leads to 
$\frac{\delta \psi}{\delta \HV_{ik}} =  \frac{\delta \psi}{\delta \P_{ik}}= 0$
and thus to $\psi = \psi(\HH_{ik})$. This is of course identical to the 
result one obtains from the reduced theory, i.e., if one starts by 
solving the constraints  classically before the transition to the quantum 
theory, but it has been obtained here following strictly the Dirac 
method, albeit in a different representation.  

Finally, we note that the classical equations of motion 
obtained from (\ref{23}) by variation with respect to $\pi^i$ and 
$\H$ have the form (on the constraint surface $\H = \pi^k = 0$)
\begin{eqnarray}\label{31}
\Delta h_{0k} + h_{0i\ ,k}^{\ \ ,i}  
- 2 \P_{,k} - \dot h^i_{\ ,i,k}
- \Delta \dot h_k &=& 0 \\ \label{32}
\Delta h_{00} - 2 \dot{\P}  &=& 0.  
\end{eqnarray}
Those equations are gauge invariant under (\ref{17}) and determine 
the multipliers $h_{00}$ and $h_{0k}$. What we wish to point out 
here is the fact that, in our specific representation 
(\ref{29}) and (\ref{30}), all the variables occurring in (\ref{31}) and 
(\ref{32}) are given in terms of multiplication operators. More generally, 
all the variables that are affected by the gauge transformations (\ref{17}) 
are given in terms of multiplication operators, meaning that the 
gauge structure of the theory can be carried over to the quantum theory 
without modification. This is to be compared with the conventional 
coordinate representation, where the transformation $\delta \pi_{ik} = 
- \epsilon_{,i,k}+ \eta_{ik} \Delta \epsilon$ does hardly make sense in 
the quantum theory. This feature of our specific representation  is not 
a coincidence, and we will come back to this point later on.

The  manipulations performed in this section were 
simple to carry out, because of the 
linearity of the theory. In particular, the projection operators 
that are used to perform the orthogonal decomposition into $TT$, $T$ and $V$ 
parts are constant, and therefore easy to handle. Things are much more 
involved in non-linear theories, for instance when we have to deal 
with {\it covariantly transverse} components or similar. In the case of 
non-abelian Yang-Mills theory, we are rather lucky, because the 
state functional turns out to be already gauge invariant in the 
coordinate representation (as a result of the linearity of the constraint 
in the momenta). In General Relativity, as we will show in the next 
section, this is not the case, 
and the representation for which the functional is gauge independent 
is very hard to find.

\section{General Relativity}

In General Relativity, the first order Lagrangian 
(in the ADM parameterization) is of the form 
\begin{equation}\label{33}
L =\int\left[ \pi^{ik} \dot g_{ik} - N \mathcal H - N_i \mathcal H^i\right]
\de ^3 x, 
\end{equation}
with 
\begin{equation}\label{34}
  \mathcal H = 
\frac{1}{\sqrt{g}}(\pi^{ik} \pi_{ik} - \frac{1}{2} \pi^2)
- \sqrt{g}\  {}^3\!R, \quad
\mathcal H^i = -2 \pi^{ik}_{\ \ ;k} = 0, 
\end{equation}
where indices are raised and lowered with the spatial metric $g_{ik}$ and 
its inverse $g^{ik}$, and the semicolon denotes covariant derivation with 
respect to $g_{ik}$ (note that $\pi^{ik}$ is a tensor density). The formal 
analysis is identical to that in the previous section. We have the four 
constraints $\mathcal H = \mathcal H^i = 0$ and thus consider the 
four gauge generators 
\begin{equation} \label{35}
G_{\epsilon} = \int \epsilon \mathcal H \de^3 x, \quad 
G_{\xi} = \int \xi_i  \mathcal H^i \de^3 x, 
\end{equation}
which arbitrary parameters $\epsilon, \xi_i$ (assumed to be such that 
surface terms can always be omitted). Those generators induce 
transformations on $\pi^{ik}$ and $g_{ik}$ 
which can be found by 
 evaluating $\frac{\delta G_{\epsilon}}{\delta \pi^{ik}} = 
\delta_{\epsilon} g_{ik}$, $\frac{\delta G_{\epsilon}}{\delta g_{ik}} = 
- \delta_{\epsilon} \pi^{ik}$ and similar for $G_{\xi}$, see (\ref{9}). 
The action of the corresponding pre-quantized operator on the pre-quantized 
gauge functional $f(g_{ik}, \pi^{ik})$ can be read off from our general 
result (\ref{11}) 
\begin{eqnarray} \label{36}
\hat G_{\epsilon} f &=& 
\frac{1}{i} \delta_{\epsilon}f - \left(\int \left[ \delta_{\epsilon} \pi^{ik} 
\theta_{ik1} + \delta_{\epsilon} g_{ik} \theta^{ik}_{\ \ 2}\right]\de^3 x
- G_{\epsilon} \right)f
\\\label{37}
\hat G_{\xi} f 
&=& \frac{1}{i} \delta_{\xi}f - \left(\int \left[ \delta_{\xi} \pi^{ik} 
\theta_{ik1} + \delta_{\xi} g_{ik} \theta^{ik}_{\ \ 2}\right]\de^3 x
-G_{\xi}\right) f 
\end{eqnarray}
In our previous examples, we used the fact that the constraints were 
linear (or, more precisely, homogenous of first degree) in certain 
variables. Here too, this is the case for $G_{\xi}$, namely 
we have again $G_{\xi} = \int (\delta_{\xi} g_{ik}) \pi^{ik}\de^3 x$ 
just as in the linear theory. Note that an analogue to the 
relation $G_{\epsilon} = 
- \int (\delta_{\epsilon} \pi^{ik}) h_{ik}\de^3 x$ of the linear theory 
does not hold here. In any case, it is easy to see that  with the 
conventional choice $\theta_{ik1} = 0$ and $\theta^{ik}_{\ \ 2} = \pi^{ik}$ 
(which  leads to $f = \psi(g_{ik})$ in the coordinate polarization), 
we find $\hat G_{\xi} f = \frac{1}{i} \delta_{\xi} f$, but 
$\hat G_{\epsilon} f \neq \frac{1}{i} \delta_{\epsilon} f$. That is, after 
imposing the constraints, the functional will be invariant under those 
transformations induced by $\hat G_{\xi}$ but not under  those induced by 
$\hat G_{\epsilon}$. 

The fact that in the coordinate representation, the effect of 
the constraints $\mathcal H^i$ is to render the wave functional 
invariant under the transformations $\delta_{\xi} g_{ik}$ and 
$\delta_{\xi} \pi^{ik}$ (i.e., under general coordinate 
transformations  in three-dimensional space) is well known, 
see \cite{higgs, dirac, dewitt}. On the other hand, the fact that 
$\hat G_{\epsilon}$ has not the effect of rendering the wave functional 
invariant under $\delta_{\epsilon} g_{ik}$ and $\delta_{\epsilon} \pi^{ik}$ 
(which correspond, on-shell,  
to the remaining spacetime diffeomorphisms, see, e.g., \cite{wald})  
has led to difficulties concerning the interpretation of this 
constraint, i.e., of the Wheeler-DeWitt equation $\mathcal H \,
\psi = 0$ (more precisely, $\hat G_{\epsilon} \psi = 0$). 
In \cite{higgs}, it is  suggested that 
$\mathcal H$ should eliminate a further degree of freedom from the 
theory, but the discussion is transferred to future work. Dirac 
\cite{dirac} claims that the meaning of the constraint is that $\psi$ 
should be independent of deformations of the surface, but this is just 
what we have proved  not to be the case. It is well-established that 
the Hamiltonian constraint $\mathcal H$ should describe the 
dynamics of the theory, see  \cite{dewitt}, but this does not really 
clarify its role as gauge generator, nor how the (remaining) 
non-physical degrees of 
freedom are to be eliminated by that same constraint. 

In fact, it 
seems as if the exact meaning of the Hamiltonian constraint 
$G_{\epsilon}\psi = 0$ is not totally understood. A characteristic 
statement is the following: {\it Roughly speaking, the constraints 
$\hat G_{\xi}\psi = 0$ and $\hat G_{\epsilon} \psi= 0$ can be 
interpreted as requiring  the invariance of $\psi$ under the infinitesimal 
canonical transformations generated by $G_{\epsilon}$ and $G_{\xi}$, 
which, as discussed above, correspond to infinitesimal coordinate 
transformations on the manifold of solutions. For the momentum constraints 
$\hat G_{\xi}  \psi= 0$, this holds literally.\ [\dots] It does not 
appear possible to give as literal an interpretation of the quantum 
constraint $\hat G_{\epsilon} \psi = 0$, known as Wheeler-DeWitt equation, 
as corresponding to the invariance of $\psi$ under a variation of $g_{ik}$ 
corresponding to an infinitesimal diffeomorphism in spacetime which moves 
points on $\Sigma$ in the direction orthogonal to $\Sigma$. Nevertheless, 
this interpretation of the quantum constraints can be viewed as accounting 
for why $\delta \psi / \delta t = 0$ in the formalism.} This statement is 
taken from an article of Unruh and Wald \cite{wald}, where for 
convenience,  we have replaced the notation with  our own. Further, 
$\Sigma$ is the three-dimensional manifold, and $\delta \psi/ \delta t = 0$ 
results because the Hamiltonian (on the constraint surface) is zero. 

While this is in accordance with our results, the question  
about the exact interpretation of the Hamiltonian constraint remains. 
On the other hand, it is clear that if one would use a representation 
in which the constraints really imply gauge invariance of $\psi$, then 
the problem would be 
 trivially solved. How this is to be achieved is clear from 
the corresponding procedure in the linear theory. Namely, instead of 
taking $\theta_{ik1} = 0$ and $\theta^{ik}_{\ \ 2} = \pi^{ik}$ (coordinate
representation), we have to modify this in in a way not to destroy the 
already obtained invariance under $G_{\xi}$. This is done by 
recognizing that not all components of $\theta^{ik}_{\ \ 2}$ are actually 
contained in the term  $\int \delta_{\xi} g_{ik} \theta^{ik}_{\ \ 2}$
in (\ref{37}). We can thus take some  components and shift them 
over (in terms of the canonically conjugated variables) to $\theta_{ik1}$, 
which results in some components being brought into the 
  momentum representation. Also, some components of 
$g_{ik}$ (and the conjugate components of $\pi^{ik}$) do not appear 
at all in (\ref{36}) and (\ref{37}), as was the case with $\HH_{ik}$ and 
$\PP^{ik}$ in the linear theory. For those, we have again the choice 
of using a coordinate of momentum representation, without the 
transformation properties of $f$ (and thus of $\psi$) being affected. 

Although this sounds all very nice, it is obvious that we cannot 
perform this explicitely in the full theory. Because, if we did, 
we would have identified the dynamical variables, and since those 
satisfy trivial equations of motion ($H = 0$), we would have solved the 
Einstein equations completely. In other words, the above procedure is 
in fact of the same degree of difficulty than solving the Einstein 
equations, or, on the quantum level,  
solving the Wheeler-DeWitt equation. There is no magic way around 
 the non-linearities of General Relativity. What one can do, however, 
is to check for specific representations, whether the corresponding 
functional is gauge invariant or not. As outlined above, this is not 
the case for the coordinate representation, and it is also not hard to 
show that the same holds true for a pure momentum representation. 

Our approach can nevertheless  be useful for the  interpretation of 
the Wheeler-DeWitt equation in another way. Namely, there is a tendency 
in the literature to relate any kind of  
difficulties one encounters during the analysis of General 
Relativity directly to the  specific features of generally covariant 
theories. Since there are indeed  certain features that one does not 
encounter in conventional theories, it would be wise to carefully isolate 
them from possible additional problems, that are not related, e.g., to the 
reparameterization invariance. In this sense, if one wishes to correctly 
interpret the Wheeler-DeWitt equation, a good starting point is to 
have a general idea of what could be the answer to 
the following questions: What does 
it  mean that a wave functional is not gauge independent  
after the  constraints have been imposed? And a directly related question: 
What is the meaning of a constraint 
if it is not to  eliminate  unphysical degrees of freedom from 
the wave functional and thus to render it invariant under the  
transformation induced by the same constraint 
 on the phase space variables?  Obviously, 
those are questions that are completely unrelated to the more 
specific problems of General Relativity concerning, e.g., the 
problem of time, despite the connection there might be 
between the later issue and  the Hamiltonian constraint. It concerns, e.g., 
the fourth constraint of the linear spin-two theory in the coordinate 
representation, but also  electrodynamics and Yang-Mills theories in 
the momentum representation.  

In section \ref{cosmo}, we will present simple models, for which  
it is possible to identify explicitely the representation(s) in which 
the constraints eliminate the gauge degrees of freedom from 
the state functional. As in the linear spin-two theory, they turn out 
to be mixed representations. Before, in section \ref{particle}, we   
briefly analyze the case of a relativistic point-particle.

\section{Relativistic point-particle} \label{particle}  

The special relativistic point-particle is described by the 
Lagrangian $L = - m  \sqrt{\dot 
x_{\mu} \dot x^{\mu}}$ ($\mu = 0,1,2,3$) 
or, in first order form,  
\begin{equation}
L = p_{\mu} \dot  x^{\mu} - \lambda(p_{\mu} p^{\mu} - m^2). 
\end{equation}
Note that we use the notation $x^0$ for the {\it time} coordinate, 
and $t$ for the parameter of the theory, i.e., $\dot x^{\mu} = \de x^{\mu}/
\de t$. The Lagrange multipliers $\lambda$ is arbitrary and leads to 
the constraint $p_{\mu} p^{\mu} - m^2 = 0$, which 
generates the transformations $\delta x^{\mu} = \epsilon \pi^{\mu}$ and 
$\delta p_{\mu} = 0$.  The constraint can  be solved  
for $p_0$ (assumed to be positive), leading directly to the reduced 
dynamics in terms of three canonical pairs of variables. Here, instead, 
we wish to perform the  quantization first and then  impose the 
constraint on the quantum  states of the theory. Note that the 
Hamiltonian itself is zero, and therefore, the Schroedinger equation 
leads to $\partial \psi/ \partial t = 0$. The dynamics, therefore, is 
obtained from the constraint alone, which leads, in the coordinate 
representation, to the Klein-Gordon equation. However, since the constraint 
is not homogenous of first degree in the momenta, it is immediately obvious 
from (\ref{11}) that in that representation (obtained again from 
$\theta_{\mu 1} = 0$ and $\theta^{\mu}_{\ 2} = p^{\mu}$), the wave 
function will not be gauge invariant, i.e., the Klein-Gordon equation 
does not remove the unphysical degrees of freedom from the theory. 
(In fact, it does not even make sense to ask the question whether 
a certain function $\psi(x^{\mu})$ is invariant under the 
transformation $\delta x^{\mu} = \epsilon p^{\mu}$, since $p^{\mu}$ 
is supposed to be represented by 
a differentiation operator. Since classically, we have $p^{\mu} = const$ 
(on-shell), one could think of considering transformations 
of the form $\delta x^{\mu} = \epsilon a^{\mu}$ with constant $a^{\mu}$ 
instead. It is not hard to show for explicit solutions of the Klein-Gordon
equation that $\psi$ will not be invariant under such transformations 
neither.)

This short-come is readily fixed by linearizing the constraint via the 
introduction of a new variable $P_0 =  p_{\mu} p^{\mu} - m^2$ and 
its conjugate $X^0 = \frac{1}{2} \frac{x^0}{p_0}$. Altogether, 
we perform the phase-space transformation 
\begin{equation}
X^0 = \frac{1}{2} \frac{x^0}{p_0}, \quad X^i = x^i - \frac{p^i x^0}{p_0}, 
\quad P_0 = p_{\mu}p^{\mu} - m^2,\quad 
\quad P_i  = p_i, 
\end{equation}
which is easily shown to represent a canonical transformation, either 
by checking the canonical  Poisson brackets or by noting that 
$p_{\mu} \dot x^{\mu} = P_{\mu} \dot X^{\mu}$ (up to a total derivative). 
The first order Lagrangian now takes the simple form ($i=1,2,3$)
\begin{equation} \label{42} 
L=  P_i \dot X^i - \lambda P_0, 
\end{equation}
where in the kinetic term, the constraint $P_0 = 0$ can be used, since 
$P_0$ is coupled with an arbitrary multiplier anyway. The constraint 
being linear in $P_0$, it is obvious that we have to put $X^0$ into 
the coordinate representation in order to obtain a gauge invariant 
wave function. Note that the only gauge variable is now $X^0$, 
transforming as $\delta X^0 = \epsilon$. The remaining variables 
$(X^i,P_i)$ are physical and  we can use any representation we 
wish. If we choose the coordinate representation, we have 
$\psi = \psi(X^i,X^0)$, and the constraint $P_0\psi
 = \frac{1}{i} \frac{\partial 
}{\partial X^0} \psi = 0$ leads to $\psi = \psi(X^i)$, which is 
gauge invariant and coincides with the result obtained by 
the explicit reduction of the theory. Although this appears now in the 
form of  
a pure coordinate representation, in terms of the initial variables, it 
is actually a mixed representation. Namely, one has to use a 
functional depending on certain combinations of the phase-space variables, 
$\psi = \psi(x^0/p_0, x^i - p^i x^0/p_0)$, and then consider the 
following operators 
\begin{eqnarray*}
\hat x^i = x^i - \frac{x^0}{p_0} (p^i - \frac{1}{i} \frac{\partial}{
\partial(x^i - p^i x^0/p_0)}),
&\quad& 
\hat p_i = \frac{1}{i} \frac{\partial }{
\partial (x^i - p^i x^0/p_0)},\\
 \quad 
\hat p_0 = \sqrt{\frac{2}{i} \frac{\partial}{\partial(x^0/p_0)} + 
\frac{\partial^2}{\partial(x^i - p^i x^0/p_0)\partial(x^i - p^i x^0/p_0)}
+ m^2}, 
&\quad &\hat x^0 = \frac{x^0}{p_0} \hat p_0. 
\end{eqnarray*}
The constraint $(\hat p_{\mu}\hat p^{\mu}-m^2) \psi = 0$ then implies  
that $\psi$ does not depend on $x^0/p_0$, and the resulting 
wave function $\psi(x^i - p^i x^0/p_0)$ is indeed gauge invariant. As one  
can see, we are actually quite polite   using the mild expression
   {\it mixed representation} for the above.  

It is further interesting to observe that the constraint equation in the new  
 variables, $\frac{\partial }{\partial X^0}\psi= 0$,  cannot be obtained 
(e.g., with $X^0$ as time variable) from a Lagrangian
(at least not without the help of auxiliary fields), quite in contrast 
to the Klein-Gordon equation. In particular therefore, we do not have 
a canonical way to perform a second quantization, i.e., to treat 
$\psi$ as a quantum operator. 

In relation to General Relativity, if we ask questions about the
interpretation of the Wheeler-DeWitt equation, we should first 
answer the corresponding questions on the interpretation of 
 the Klein-Gordon equation. Apart from the fact that it 
describes (in the coordinate representation) the dynamics of the system, 
we would like to know its relation to the symmetry 
transformations and to the elimination of the unphysical degrees of freedom. 
All we can give here is a negative statement: The Klein-Gordon equation 
 does not remove the gauge degrees of freedom from the wave function.    
On the other hand, in the new representation, the interpretation of 
the constraint equation is completely clear. It removes the gauge 
variable from the wave function, which is also all the dynamics we 
get. For the rest, the explicit form of $\psi$ has to be determined 
by physical arguments (boundary conditions).

\section{Cosmological models} \label{cosmo}

First, we consider homogenous spacetimes with metric 
\begin{equation}
\de s^2 = N^2 \de t^2 - R^2\left( \exp(2 b_1) \de x^2 + \exp(2 b_2) \de y^2 
+ \exp(-2 b_1 - 2 b_2) \de z^2\right), 
\end{equation}
depending on 4 variables $N(t), R(t), b_1(t)$ and $b_2(t)$. The 
corresponding vacuum Einstein field equations are derived from 
\begin{equation}
L = \frac{1}{N} [ - 2 R^3 (\dot b_1^{\, 2} + \dot b_1 \dot b_2 + 
\dot b_2^{\, 2}) 
+ 6 R \dot R^2 ]. 
\end{equation}
Such models have been studied in \cite{misner} where the detailed analysis 
can be found. It is convenient to perform a change of variables, defining 
\begin{equation} \label{45}
\tilde N = \frac{N}{2} \exp(- \sqrt{3} X), \quad X =  \sqrt{3}
\ln R. 
\end{equation}
This leads to $L = \frac{1}{\tilde N}(- (\dot b_1^{\,2} + \dot b_1 \dot b_2 + 
\dot b_2^{\,2}) + \dot X^2)$ and to the first order Lagrangian 
\begin{equation}
L = \pi \dot X + p_1 \dot b_1 + p_2 \dot b_2 - \tilde N \left[
\frac{\pi^2}{4} - \frac{1}{3} (p_1^{\,2} + p_2^{\,2} - p_1 p_2)\right], 
\end{equation}
where $\pi, p_1$ and $p_2$ are the momenta conjugated to $X, b_1$ and $b_2$, 
respectively. One can directly solve the constraint for $\pi$ and 
derive the reduced Lagrangian. Here instead,  we are interested again in 
imposing the constraint on the quantum states. Since the constraint 
is not homogenous of first degree in neither variable, we perform a 
canonical transformation 
\begin{eqnarray}\label{47}
B_1 = b_1 + \frac{4}{3}\ \frac{ p_1 - \frac{1}{2} p_2}{ \pi} \ X, \quad 
B_2 = b_2 + \frac{4}{3}\ \frac{ p_2 - \frac{1}{2} p_1}{ \pi} \ X, \quad \\
Y = \frac{2 X}{ \pi}, \quad \Pi = \frac{\pi^2}{4}
 - \frac{1}{3} (p_1^2 + p_2^2 - p_1 p_2),  \label{48}
\end{eqnarray}
with $p_1$ and $p_2$ as before. We now find 
\begin{equation}\label{49}
L = p_1 \dot B_1 + p_2 \dot B_2 + \Pi \dot Y -  \tilde N \Pi.   
\end{equation}
The fact that the kinetic terms are still in canonical form (and thus 
the symplectic two-form is unchanged) proves that the above transformation 
was indeed canonical. The constraint $\Pi = 0$ induces the gauge 
transformation $\delta Y = \epsilon$, under which $L$ is invariant, 
if we assume in addition $\delta \tilde N = \dot \epsilon$.  
Since the constraint is  linear in $\Pi$, 
  a gauge invariant 
wave functional is obtained by choosing a coordinate representation 
for $Y$, i.e., $\hat Y   = Y$ 
and $\hat \Pi = \frac{1}{i} \frac{\delta }{\delta Y}$ The variables 
 $B_1$ and $B_2$ are physical and can be taken in any representation. 
They represent the two dynamical degrees of freedom of the (homogenous) 
gravitational field. In particular, in the coordinate representation, 
we have $\psi = \psi(B_1, B_2, Y)$, 
and the constraint imposed on  $\psi$ eliminates the gauge 
variable $Y$, i.e., $\psi = \psi(B_1, B_2)$, which is trivially 
gauge invariant and is identical to the results one obtains by direct 
elimination of $\Pi$ from the classical theory  (reduction).  
Similar as in the case of the point-particle, 
the representation, when expressed in terms of the initial variables, 
is far from being a pure coordinate representation.  We also note that 
from (\ref{49}), we find the equation of motion $\tilde N - \dot Y = 0$, 
which determines $\tilde N$. The equation is gauge invariant, and 
we observe further that $\tilde N$ (and thus also $N$, see (\ref{45}) and 
(\ref{48})) is given, in the specific representation, in terms 
of simple multiplication operators.  

As a further example, we consider flat Robertson-Walker spacetimes, 
$\de s^2 = N(t)^2 \de t^2 - R(t)^2 \delta_{ik} \de x^i \de x^k$, 
  with a minimally coupled massless scalar field $\phi(t)$. 
The Lagrangian is here (see \cite{isham,leclerc2})
$L = - \frac{6R}{N} \dot R^2 + \frac{R^3}{2N} \dot \phi^2$ which leads 
to the first order Lagrangian $L = \pi_R \dot R + \pi_{\phi} \dot \phi + N 
(\frac{\pi_R^2}{24 R} -\frac{\pi_{\phi}^2}{2 R^3})$. Performing a change 
of variables $X = \phi - \sqrt{12} \ln R,\  Y = \phi + \sqrt{12} \ln R$ and 
$\tilde N = 2 R^{-3} N$ in the second order Lagrangian, we obtain the  
simplified first order form 
 \begin{equation} \label{50}
L = \pi_X \dot X + \pi_Y \dot Y + \tilde N \pi_X \pi_Y, 
\end{equation}
with constraint $\pi_X \pi_Y = 0$, which, 
in contrast to the initial form, is free of ordering problems. The induced 
transformations are $\delta X = -\epsilon \pi_Y$, $\delta Y = -\epsilon \pi_X$
and $\delta \pi_X = \delta \pi_Y = 0$. The Lagrangian is invariant if 
$\delta \tilde N = \dot \epsilon$. 
Classically, we can solve the 
constraint by $\pi_X = 0$ or by $\pi_Y = 0$, resulting in 
$\delta Y = 0$ or $\delta X = 0$. Thus, depending on the set of 
classical solutions one refers to, the dynamical variables are either 
the pair $(X, \pi_X)$, or the pair $(Y, \pi_Y)$. On the other hand, 
the Wheeler-DeWitt equation, in the conventional coordinate representation, 
leads to $\frac{\partial^2}{\partial X \partial Y}\psi(X,Y) = 0$, 
i.e., $\psi = \psi_1(X) + \psi_2(Y) $, which is not gauge invariant in 
neither case. We conclude that the Wheeler-DeWitt equation does not 
eliminate the unphysical degrees of freedom from the state functional. 
This is consistent with the general relation (\ref{11}) and the fact 
that the constraint is not of first degree in the momenta. 

On the other hand, the constraint is linear in each momentum $\pi_X$ and 
$\pi_Y$ separately. Therefore, if we choose a mixed representation 
with either $X$ or $Y$ in the momentum representation, the resulting 
wave functional should be gauge invariant. Let us therefore adopt 
the following representation
\begin{equation} \label{51}
\hat X = - \frac{1}{i} \frac{ \partial}{\partial \pi_X},\quad \hat \pi_X = 
\pi_X, \quad \hat Y = Y, \quad   \hat \pi_Y = \frac{1}{i} \frac{\partial}{
\partial Y},  
\end{equation}
and $\psi = \psi(\pi_X, Y)$. 
The constraint equation now reads 
$\left(\pi_X \frac{\partial }{\partial Y}\right) \psi(\pi_X,Y) = 0$. 
There are two cases 
to consider. First, if $\pi_X$ is different form zero, then the solution 
of the constraint is $\psi= \psi(\pi_X)$ which is gauge invariant in 
view of $\delta \pi_X = 0$. On the other hand, if $\pi_X = 0$, then 
there is no condition on $\psi$, which is therefore a general function of 
$Y$, $\psi = \psi(Y)$. This too is gauge invariant, since $\delta Y 
=  - \pi_X = 0$. Nevertheless, we would rather consider this as 
a condition on $\pi_X$ and not on $\psi$ (similar to the constraint 
$\Delta \H\, \psi  = 0$ in the spin-two theory) and conclude that in this 
case, we are not using the appropriate representation. We then choose instead 
the representation with the roles of $X$ and $Y$ interchanged and end up 
with $\psi = \psi(\pi_Y)$.  The fact that we have to distinguish 
between two cases 
is typical for General Relativity, where the dynamical variables cannot 
be determined in a general form (e.g., in terms of traceless-transverse 
components of $g_{ik}$ or similar), but depend on the specific form of 
the corresponding classical solutions. This is already obvious from the 
fact that the constraint $\pi_X \pi_Y = 0$ has two sets of 
classical solutions. See also the discussion in \cite{leclerc2}. 

Thus, once again, a mixed momentum/coordinate representation is needed in
order for the constraint to do what it is expected to do, namely to eliminate 
the unphysical degrees of freedom from the theory. It is not hard to verify 
that (\ref{51}) corresponds to a mixed representation also in the initial 
variables $R,  \pi_R, \phi, \pi_{\phi}$. 

We also note that from (\ref{50}), we find the gauge invariant equations
$\dot X + \tilde N \pi_Y = 0$ and $\dot Y + \tilde N \pi_X = 0$. Thus,
if $\pi_X \neq 0$, we can express $\tilde N$ in terms of the multiplication 
operator $Y$, while in the other case, we take again the second 
representation and express $\tilde N$ in terms of the multiplication operator 
$X$. On the other hand, in the conventional coordinate representation
(Wheeler-DeWitt approach), $\tilde N $ is represented by a 
differentiation operator, see \cite{leclerc2}.  

\section{Discussion}

The method for obtaining a gauge invariant state functional is now 
clear. It consists simply in using a representation where 
the {\it gauge variables} are represented by multiplication operators, 
and the corresponding, canonically conjugated {\it constrained variables} 
by differentiation operators. In that way, the state functional 
will depend on the gauge variables, which are then eliminated by the action of 
the constraints. For the physical variables, there is no restriction, and 
we have a complete symmetry between momentum and coordinate representation. 
The geometric quantization formalism presented in section \ref{geo} is simply 
a justification for this simple recipe, which one might or might not find 
useful. 

There remains the question of the meaning of the constraints when we use 
a representation different from the above. Consider, for instance, 
the forth constraint of the spin-two theory, $\Delta \H = 0$. If 
we use a pure coordinate representation, $\psi = \psi(\HH_{ik}, h_i, \H)$, 
then the remaining three constraints eliminate the gauge variable $h_i$. 
Thus, we are left with $\psi= \psi(\HH_{ik}, \H)$. But this is already 
gauge invariant. This shows once again that the effect of the fourth 
constraint cannot be to render $\psi$ gauge invariant.  Instead, 
we have argued that the equation $\Delta \H \psi(\HH_{ik}, \H) = 0$ 
(suppressing the integration and the gauge parameter for simplicity) is 
actually not a condition on $\psi$, but rather on $\H$, and leads to 
$\Delta \H = 0$ (or $\H = 0$). Since this is rather an explicit 
elimination (reduction) of variables via a classical equation of motion 
than the solution to  a quantum constraint equation, we considered this 
to be outside of the spirit of the original Dirac method. There is, however, 
a different interpretation to the above equation, which is adopted, e.g., 
in \cite{jackiw2}, namely $\Delta \H \psi(\HH_{ik}, \H) = 0$ is to 
be solved by functionals $\psi$ with support only on configurations 
with $\Delta \H = 0$ (or $\H = 0$). While physically, this 
leads to the same results, from a mathematical standpoint, it is a quite 
different statement than merely requiring $\Delta \H = 0$, since it is 
indeed a condition on $\psi$ now. Nevertheless, we are still 
not completely happy with such an interpretation, because it does not 
really render $\psi $ independent of $\H$ (as is the case in the mixed 
representation). Since, if $\psi$ were independent of $\H$, there would 
obviously be no need to differentiate between $\H \neq 0$ and $\H  = 0$. 
The functional still depends implicitly (via the definition of the support) 
on $\H$. A similar situation holds in electrodynamics, if we use a pure 
momentum representation, $\psi =\psi(E)$. Also here, the functional 
is already gauge invariant (since $\delta E^i = 0$), and one can 
interpret the constraint $E^i_{\ ,i} \psi = 0$ by requiring that $\psi$ 
has support only on the transverse part of $E^i$. 

While the above interpretation works quite well and differs only 
esthetically from the corresponding approach with mixed representations 
(after all, we are not really interested in the gauge dependent parts 
anyway, so any way of eliminating them should be equally acceptable), 
the situation is different for more complicated theories. Consider 
non-abelian Yang-Mills theory in a pure momentum representation, with 
$\psi =\psi(E)$. This is not gauge invariant, and it is 
also not gauge invariant after the constraint has been imposed, as we have 
shown in section \ref{geo}. Recall that the  constraint reads   
$G 
= \int \de^3 x\ \epsilon^a (\partial_i E^i_a + f_{ab}^{\ \ c} A^b_i E^i_c)$. 
But now, the equation $\hat G \psi(E)$ cannot be interpreted 
as {\it `` $\psi(E)$ has support only on configurations with 
$\partial_i E^i_a + f_{ab}^{\ \ c} A^b_i E^i_c= 0$''}, since the  
latter expression contains the operator $A^b_i =  - \frac{1}{i} \frac{\delta}
{\delta E^i_a}$. An idea is to treat the above partly as a condition 
on the support of $\psi$ (as in electrodynamics) and partly as 
differential equation. Namely, we first imply the second term 
of the constraint on $\psi$, in the form $\int \de^3 x\  
\epsilon^a f_{ab}^{\ \ c} E^i_c 
\frac{\delta }{\delta E^i_b} \psi(E) = 0$. It is not hard to show that 
this requires $\psi(E)$ to be gauge invariant (since $E^i_a$ transforms 
covariantly, i.e., as a vector). In addition, we interpret the remaining 
term of the constraint as in electrodynamics, namely that $\psi(E)$ 
has only support on the transverse part of $E^i_a$ (i.e., on $E^i_a$ 
satisfying $E^i_{a ,i} = 0$).  In this way, we end up with gauge invariant 
functionals $\psi(\EE)$, e.g., $\psi = \psi(\EE^i_a \EE^{ak})$.  
 The only thing we have 
to worry about is whether  we have possibly imposed too  many  conditions on 
$\psi$, because it seems to be a stronger requirement for each term of 
the constraint separately to annihilate $\psi$ than merely to impose the  
constraint. 
This is true (since we have already shown that $\hat G \psi= 0$ alone 
does not 
lead to a gauge invariant $\psi$), but it is not physically relevant. 
The gauge dependent parts of $\psi$ are unphysical anyway, so rendering 
$\psi$ gauge invariant cannot do any harm. Once this has been done, the 
remaining condition $E^i_{a ,i} \psi =0$ is a direct consequence. So, 
in a sense, we have done more than the constraint requires, but this
concerns only unphysical contributions in $\psi$. Note that the above 
argumentation is not unsimilar to the approach adopted in  \cite{jackiw} and 
\cite{jackiw2}, where a gauge dependent factor is extracted from the 
wave functional, leading to a functional that is annihilated by the 
rotational part of the constraint generator. On the other hand, it is 
hard to imagine how a similar argumentation could be applied in the case of 
the Wheeler-DeWitt equation.  (Also, it fails completely in the case of the 
point-particle in the coordinate representation.)   

Let us compare the above  with the approach presented in section \ref{geo}, 
which is based on the use of a gauge invariant functional.  
Being  interested in a momentum representation, we should, at least, 
put  the dynamical variables into the momentum representation. Even though 
 we do not know exactly which are the physical  (gauge invariant) 
components of $A^a_i$, we can recognize that  the 
transverse components $\AA^a_i$ 
transform covariantly  (i.e., as vectors). (This holds actually only 
for infinitesimal transformations, see below.)
Therefore, let us choose 
$\theta^i_{a2} =  \EL^i_a$ and $\theta^a_{i1} = - \AA^a_i$. (Our 
notation is $B_i = {}^{\text{\tiny T}}\!B_i + {}^{\text{\tiny L}}\!B_i$, 
with ${}^{\text{\tiny L}}\!B_i = \frac{\partial_i \partial^k}{\Delta}B_k$.) 
 From 
(\ref{11}), we find that $\hat G = \frac{1}{i} \delta f$, as required. 
We use the polarization ${}^{\text{\tiny T}}\!\De^i_{a2} f = 0$ and  
${}^{\text{\tiny L}}\!\De_{i1}^a f = 0$, leading to $\delta f/ \delta \AA = 0$ 
and $\delta f /\delta \EL = 0$, i.e., to $f = \psi(\AL, \EE)$.  
The constraint now leads to 
\begin{equation} \label{52}
\hat G \psi = \frac{1}{i} 
\int \de^3 x \left(\frac{\delta \psi}{\delta \AL^a_i}(-\epsilon^a_{\ ,i} 
- f_{db}^{\ \ a}\ \AL^b_i \epsilon^d) - \frac{\delta \psi}{\delta \EE^i_a}
f_{da}^{\ \ c} \ \EE^i_c \epsilon^d \right) = 0, 
\end{equation}
which results in
$\delta \psi /\delta \AL^a_i = 0$ and $\int \de^3 x 
\frac{\delta \psi}{\delta \EE^i_a}
f_{da}^{\ \ c} \ \EE^i_c = 0$. The second relation is again recognized as 
the invariance of $\psi(\EE^i_a)$ under rotations, and thus, we find 
the same results as before. Thus, if we start directly from a 
gauge invariant pre-quantized functional $f$, there is not need to 
split the constraint into two terms or to extract a gauge dependent 
phase from the functional. One is directly led to a consistent momentum 
representation. 

We conclude that, although in certain cases, 
it is possible to interpret constraints which 
contain variables in form of multiplication operators as a condition 
on the support of the wave functional, such a procedure is, on one hand, not 
really elegant, because it means that implicitly the wave functional 
depends nevertheless on those variables, and on the other hand, it leads to 
additional difficulties in more complex cases, where one  has to 
interpret  a part of the constraint as differential equation and another 
 part as condition on the support. In other cases, for instance 
the relativistic point-particle in the coordinate representation, there 
does not seem to exist a similar way to interpret the constraint.   
It is therefore more systematic and more straightforward to 
choose a representation that eliminates all unphysical fields 
from the wave functional directly by true differential equations. 

We should note, however, that the above argumentation was based on 
infinitesimal gauge transformations and does not take account  of 
the  problems related to the Gribov ambiguity. In fact, it 
is not completely accurate to consider $\epsilon^a$ and $\epsilon^a_{\ ,i}$ 
in (\ref{52}) as independent. There are functionals that depend on 
$\AL^a_i$ and for which the above $\hat G \psi$ is zero nevertheless. 
For instance, we can take $\psi = \int F^a_{\ ik}(\AL) F_a^{\ ik}(\AL) 
\de^3 x$, where $F^a_{\ ik}(\AL)$ is the Yang-Mills tensor with $\AA^a_i =
0$. This leads to $\hat G \psi = 0$, because we have $\delta \psi /\delta \AL 
\sim \De_i F^{ik}$ and $\De_i \De_k  F^{ik} = 0$, where both $F$ and the 
covariant derivative $\De $ are formed with $A = \AL$. Thus, $\De (\delta
\psi/\delta \AL) = 0$, which corresponds to 
 the two first terms in (\ref{52}). 
The problem is related to the fact that $\AA^a_i$ and $E^i_a$ do not 
describe the theory completely, since, e.g.,  if $E^i_a = \AA^a_i = 0$, 
 the gauge invariant quantity $F^a_{\ ik} F_a^{\ ik}$ 
can still be different from zero, and can be further traced back to the 
fact that the decomposition $A = \AA + \AL $ is not gauge invariant.  
Therefore, the gauge invariant representations do not  replace the 
detailed analysis carried out, e.g.,  in \cite{jackiw,baluni}, 
concerning the identification of the physical variables, 
but they could nevertheless provide an alternative starting point 
which deserves further study.

As to General Relativity, it  
represents a  special case. As we have mentioned 
before, here the dynamical variables cannot be specified in a 
general form. This is a matter of principle, and not a result of 
us being unable to solve the corresponding equations. This means in 
particular that we cannot represent  the gauge variables by multiplication 
operators and the constrained variables by differentiating operators 
without specifying a specific set of classical solutions we refer to. 
We have encountered 
this feature in our second example in section \ref{cosmo}, where two 
different representations had to be used, according to whether 
$(X,\pi_X)$ or $(Y, \pi_Y)$ are the physical variables. This is completely  
similar to the problems one faces during the explicit reduction of 
the theory, where one has to solve the classical equation $\pi_X \pi_Y = 0$ 
either by $\pi_X =0$ or by $\pi_Y = 0$, see \cite{leclerc2}, or the 
corresponding problems that arise when one decides to fix the gauge 
explicitely (choice of time coordinate), which has to be done by 
$X = f(t) $ in one case, and by $Y = f(t)$ in the other. One could 
see in this an advantage of the conventional Wheeler-DeWitt approach 
(coordinate representation), where this question is left open. There, we 
have  solutions $\psi(X,Y)
= \psi_1(X) + \psi_2(Y)$ and we do not need to decide 
whether $X$ or $Y$ is the physical variable. While this seems reasonable 
from the point of view that the quantum theory should not directly 
refer to a specific set of classical configurations (e.g., $\pi_X = 0$ or 
$\pi_Y = 0$), it is nevertheless also clear that sooner or later,   
one has to remove the gauge dependent parts from $\psi(X,Y)$. 
Whether one does this directly as conditions on $\psi$ or via boundary 
conditions or by  an appropriate construction of the  scalar product, it 
does not really change the fact that, in order to remove them, 
one will first have to decide which degrees are physical and which are not. 
In any case, we retain the fact that, 
while  the exact meaning of the Wheeler-DeWitt equation 
in the coordinate representation 
is unclear \cite{wald}, our analysis shows that there are representations 
for which there is no doubt whatsoever on the physical interpretation 
of  the constraint. 

There is one more important feature of the gauge invariant representations. 
As we have mentioned at the end of section \ref{spin2}, 
in the spin-two theory, we found that the 
Lagrange multipliers $h_{00}$ and $h_{0i}$ could be expressed in terms of 
multiplication operators. The same was shown for  $N$ (or $\tilde N$)
 in our cosmological 
examples, section  \ref{cosmo}, 
and it is not hard to argue  that this will be the case for 
Lagrange multipliers in general, whenever we use our specific
representations. (That results because, for the wave functional to 
become gauge invariant,  the constraints the multipliers 
 apply to must be homogenous of  first degree in the  variables
which are represented by differentiation operators.) While this is in 
most cases not of interest, since the multipliers are unphysical anyway, 
we have shown in \cite{leclerc2} that the situation is different 
in the linear spin-two theory and in General Relativity. Namely, while 
in conventional theories, the configuration can be completely 
specified by the constrained variables and the dynamical variables alone, 
in the latter cases, there are physical (i.e., invariant) quantities 
that cannot be expressed without the help of the multipliers. This concerns 
for instance the (four dimensional) scalar curvature in General Relativity
which contains $N$. More specifically, the equations (\ref{31}) and (\ref{32}) 
in the spin-two theory as well as equation $\tilde N - \dot Y = 0$ 
in our first example in section \ref{cosmo} and the equations 
$\dot X + \tilde N \pi_Y = 0$ and  $\dot Y + \tilde N \pi_X = 0$ 
 in the second example of the same section are all gauge invariant.
(In the latter case, only one of both will determine  $\tilde N$, according to 
whether $\pi_X = 0$ or $\pi_Y = 0$.) 
It is clear, at least on the classical level, that one cannot do 
without those equations, since if they are omitted, this means that 
 even after the gauge is fixed, there remains one undetermined invariant 
expression in the theory. More explicitely, in a Lagrangian  of type 
(\ref{49}), we would usually omit the third term, arguing that $\Pi$ 
is coupled with an arbitrary multiplier anyway (fourth term), just as 
we did in (\ref{42}) with the term $P_0 \dot X^0$.  
This, however,  would lead to $\tilde 
N = 0$ (and thus $N = 0$), which is obviously unacceptable. 

On the other hand, the equations for the multipliers $N, N_i, h_{00}, \dots$ 
can only be 
obtained by classical means, namely by variation of the Lagrangian
{\it before} the constraints are solved. Thus, such equations are neither 
quantum constraints to be imposed  on the states, nor are they obtained, 
e.g., in the form of  Heisenberg equations, as are the dynamical equations. 
In fact, 
they are irrelevant for the quantum dynamical evolution of the system and do 
not appear, e.g., in the Schroedinger equation.  
 Thus, as has been outlined in 
\cite{dewitt}, they simply represent operator relations, that is, 
$N, N_i, h_{00}\dots$ are given in terms of the remaining operators. 
(Although irrelevant for the true quantum dynamics, they can still not be 
omitted, since after all, the quantum theory should also contain  the 
classical limit.) Thus, we have the rather strange situation that we 
have quantum operators that can only be  determined by a  classical 
variation. In the conventional,  pure coordinate representation, 
it turns out that $h_{00}, h_{0i}$ as well as $N$ in cosmological models 
\cite{leclerc2} are given in terms of differentiation operators 
(and their inverse). Using the  invariant representations instead, 
they are all given in terms of simple multiplication operators. In other
words, they can be treated as classical functions. This seems much more 
appropriate for a quantity that has to be determined classically from 
a Lagrangian, and we consider this an additional advantage of the specific 
representations. Possibly, this observation 
 could  also be useful in the reversed 
direction, namely for the determination of  the invariant representations. 
That is, instead of 
analyzing directly the constraints, one could analyze the field 
equations for $N$ and $N_i$ and choose the representation such that 
$N$ and $N_i$ are given in terms of classical functions. 

Moreover, the fact 
that both the gauge variables as well as the multipliers are represented by 
multiplication operators in our specific representations allow us to carry
over the gauge structure of the theory directly to the quantum case without 
modification. As mentioned before, e.g., the transformation $\delta \pi_{ik} 
= \epsilon_{,i,k} + \eta_{ik} \Delta \epsilon $ would be hard to interpret 
in the quantum theory if we use a pure coordinate representation. As is easily
checked, we encounter  the same situation in our  remaining examples. 
In this sense, the use of our specific  representations 
combines both the features of 
the Hamiltonian reduction method (Faddeev-Jackiw), 
where only the dynamical degrees of freedom 
are quantized right from the start, and the conventional Dirac procedure, 
where we quantize the complete set of canonical variables and impose the 
constraints on the physical states. Namely, although we follow 
 the Dirac procedure, at the end, the Lagrange multipliers and the gauge 
variables are given by multiplication operators anyway, 
just as if they had not been quantized at all.

\section{Summary}

It is conventionally assumed that the constraints arising in gauge theories 
play, in the quantum theory, the role of eliminating the unphysical degrees 
of freedom from the state functional. We have shown in this article that 
for this to hold, one has to use very specific operator  representations. 
Performing an analysis based on the approach of geometric quantization, 
it was shown that for the constraints to render the state functional 
gauge invariant, the {\it constrained} variables have to be represented by 
differentiation operators, while the corresponding canonically 
conjugated  {\it gauge}  variables are to be represented by multiplication
operators. No restriction exists for the representation of the 
physical (dynamical) variables. 

The conventional coordinate representation of electrodynamics 
and non-abelian Yang-Mills theory satisfy the above requirements, but not, 
e.g., the  pure momentum representation. On the other hand, in the linear 
spin-two theory and in General Relativity, the coordinate representation 
is easily shown not to be of the required  form. 
This results, e.g., in the fact 
that the Wheeler-DeWitt equation does not render the wave functional 
invariant under variations of the metric that correspond to infinitesimal 
diffeomorphisms orthogonal to the three-dimensional space manifold. 
In the linear theory, it is readily shown that gauge invariance of the 
functional can only be achieved with representations where certain 
components of $h_{ik}$ are represented by multiplication operators, and 
others by differentiation operators, i.e., with  mixed momentum/coordinate 
representations. In  General Relativity, the corresponding 
representations cannot be found in general, but we have obtained 
similar results for simple cosmological mini-superspace models. 
In contrast to the 
 Wheeler-DeWitt equation in the conventional 
coordinate representation, whose physical meaning in relation 
to the elimination of unphysical degrees of freedom is unclear, 
in the new, mixed representations, the interpretation is straightforward, 
since the constraint now explicitely removes the gauge variables from 
the state functional, leaving us with the physical, gauge invariant 
functional.


\begin{thebibliography}{12}

\bibitem{jackiw} J.~Goldstone and R.~Jackiw, Phys.\ Lett. {\bf 74B}, 81 
(1978)

\bibitem{jackiw2} R.~Jackiw, hep-th/9604040 (1996) 

\bibitem{faddeev} L.~Faddeev and R.~Jackiw,
Phys.\ Rev.\ Lett. {\bf 60}, 1692 (1988)

\bibitem{adm} R.~Arnowitt, S.~Deser, and C.~W.~Misner,
in {\it Gravitation: an introduction to current research}, L. Witten,  ed.
(Wiley, New York, 1962), p227-264, (available as: gr-qc/0405109)

\bibitem{leclerc} M.~Leclerc,  gr-qc/0612125 (2006)

\bibitem{leclerc2} M.~Leclerc, gr-qc/0702077 (2007)

\bibitem{higgs} P.~W.~Higgs, Phys.\ Rev.\ Lett. {\bf 1}, 373  (1958) 

\bibitem{dirac} P.~A.~M.~Dirac, Phys.\ Rev. {\bf 114}, 924 (1959) 

\bibitem{dewitt}  B.~S.~DeWitt,  Phys.\ Rev. {\bf 160},  1113  (1967)

\bibitem{wald} W.~G.~Unruh and R.~M.~Wald, 
Phys.\ Rev.\ D{\bf 40}, 2598 (1989) 

\bibitem{misner} C.~W.~Misner, Phys.\ Rev. {\bf 186}, 1319 (1969)

\bibitem{isham} W.~F.~Blyth and C.~J.~Isham,
Phys. Rev. D {\bf 11}, 768 (1975)

\bibitem{baluni} V.~Baluni and B.~Grossman, Phys.\ Lett. {\bf 78B}, 226 (1978)

\end{thebibliography}
\end{document}